\documentstyle[12pt]{article}
\input dspace12
\def\be{\begin{equation}}
\def\ee{\end{equation}}
\def\bearr{\begin{eqnarray}}
\def\bearrs{\begin{eqnarray*}}
\def\eearr{\end{eqnarray}}
\def\eearrs{\end{eqnarray*}}
\def\barr{\begin{array}}
\def\earr{\end{array}}

\def\l{\left}
\def\r{\right}

\def\o{\over}

\def\non{\nonumber}

\title{\large \bf Leptonic flavor violations\\ in the presence of
an extra $Z$ }
\author{Gautam Dutta, Anjan S. Joshipura and K.B. Vijaykumar\\
Theory Group, Physical Research Laboratory\\
Navrangpura, Ahmedabad 380 009, India}
\date{}
\maketitle
\begin{abstract}
Gauged extensions [$SU(2)_L\otimes U(1)_Y\otimes U(1)_X$]
of the standard  $SU(2)_L\otimes U(1)_Y$ model obtained without
extending the fermion content of the model are studied.
Models that are possible when $U(1)_X$ is identified with some
combination of the family lepton numbers are systematically
classified.
Most of these contain flavor violations in the leptonic sector.
These flavor violations are correlated to the mixing of the
$U(1)_X$ gauge boson $Z'$ with the ordinary $Z$ in the models
considered here. Detailed phenomenological implications of a
typical model are discussed. Constraints on the $Z'$ mass and
the $Z$-$Z'$ mixing following from ({\em i}) the observations at LEP,
({\em ii}) rare processes like $Z\rightarrow e\tau$ and ({\em
iii}) flavor violating
$\tau$ decays are presented. It is found that the constraints
coming from the LEP allows the rare processes like $\tau
\rightarrow eee $ at the level of the present limits on its
branching ratio. Thus the future search could either improve on the
existing LEP limits or would find such flavor violating decays.
\end{abstract}
\newpage

\begin{document}
%\doublespace
\middlespace
\vspace{0.5 in}

\noindent

\section{Introduction}
\noindent Lepton number $L_i$ for each family $i=e,\;\mu,\;\tau$ are
globally conserved in the standard $SU(2)_L \otimes U(1)_Y$ model.
Within the minimal $SU(2)_L \otimes U(1)_Y$ theory, these symmetries
have to
be global since attempt to gauge them would introduce anomalies
and would spoil the renormalizabilty of the theory. Nevertheless,
it is possible to gauge some linear combinations of $L_i$.
Specifically, it was shown in ref. \cite{He} that only three linear
combinations of $L_i$ namely $X=L_e-L_\mu,\; L_e-L_\tau$ and
$L_\mu-L_\tau$ are gaugable and only one can be gauged along with
$SU(2)_L \otimes U(1)_Y$ at a time. Thus the maximal permissible gauge
group with the minimal content of fermions and higgs field is
$SU(2)_L \otimes U(1)_Y \otimes U(1)_X$. One could easily enlarge the
gauge symmetry
by adding more fermions as happens for example in the left-right
symmetric or $E_6$ models \cite{E6}. But it is possible to enlarge
the
available choices of $U(1)_X$ by adding more Higgs doublets
transforming non-trivially under it. The point is that the
requirement of anomaly cancellations allow for  more than the
above mentioned three possibilities for $X$. In the absence of Higgs
doublets which are non-trivial under $U(1)_{X}$, these additional
choices of $U(1)_{X}$ are physically indistinguishable from the
above three choices. Otherwise, they are inequivalent and could
lead to different predictions. $U(1)_X$ groups of this type fall
in the category of horizontal symmetries characterised by
enlargement of the gauge and Higgs sector. Many examples of such
gauge symmetries have been proposed \cite{Weinberg} and studied.
Crucial tests of such symmetries are flavor violations
associated with these symmetries. Most study of these flavor
violation were done before the results of the $e^+e^-$ collider
became available. These results can provide additional
constraints on such theory. We wish to study here simplest of
such horizontal symmetries. We shall study various choices of
$U(1)_X$ under the assumption  that ({\em a})
the fermion sector of the standard model is not extended,
({\em b}) $X$ is some linear combination of lepton family numbers,
({\em c}) the Higgs sector of the SM is enlarged by adding one or
more Higgs doublets transforming non-trivially under the gauge group.

It is possible to systematically classify all the allowed choices of
$U(1)_X$ in the presence of an enlarged Higgs sector.
We do such a classification. Different $U(1)_X$ groups studied
here differ from groups of ref. \cite{He} in three important ways.
The
$U(1)_X$ current in the present case is non-vectorial. Secondly,
the $U(1)_X$ current due to its horizontal nature leads to
flavor violations in the leptonic sector. Thirdly, the neutral
gauge boson $Z'$ associated with $U(1)_X$ necessarily mix with $Z$
in these models. The last two properties put significant
constraints on parameters of the model. We systematically work
them out. It follows from the analysis presented here that the
observable flavor
violations, e.g. in $\tau \rightarrow eee$ decay are possible within
the models in spite of the severe constraints imposed by the LEP
observations.

We shall discuss in the next section all possible choices of
$SU(2)_L \otimes U(1)_Y \otimes U(1)_X$ and general structure of the
current coupled
to $Z$ and $Z'$. Then we discuss a specific model in section 3.
Section 4 contains a discussion on constraints on parameters
taking the model of section 3 as an illustration. A summary is
contained in section 5.

\section{Possible Extensions}
  We shall confine ourselves to the minimal fermionic content as in
the standard model but would consider a general gauge group
$SU(2)_L \otimes U(1)_Y \otimes U(1)_X$, where $X$ is taken to be a
linear combination of three lepton numbers. In general $X$ need
not act vectorially on the weak interaction basis $e'_{i_{L,R}}
(i=1,2,3)$
although the $X$ assignments of members of a given $SU(2)_L$ doublet
have to be identical. For notational convenience let us write
$X$-charges in terms of diagonal matrices in the generation space.

\[
X_{L,R}={\hbox {diag}}(\alpha_1,\alpha_2,\alpha_3)_{L,R}
\]

\noindent $X_L$ determine the $U(1)_X$ assignment of the
leptonic doublet while $X_R$ that of the charged right-handed
leptons. The possible choices of $\alpha_{iL}$ and $\alpha
_{iR}$ are restricted due to anomaly cancellation which require:
\be
\barr{rcrcl}
\sum_{i}\alpha_{iL}&=&\sum_{i}\alpha_{iR}&=&0\\
\sum_{i}\alpha_{iL}^{2}&=&\sum_{i}\alpha_{iR}^{2}& &\\
2\sum_{i}\alpha_{iL}^{3}&=&\sum_{i}\alpha_{iR}^{3}& &\\
\earr \label{f1}
\ee

These constraints can be satisfied by taking any two of $\alpha_{iL}$
 and $\alpha_{iR}$ to be $\pm 1$ and the third to be zero. This can
be
done in variety of ways but a particularly simple choice results
when one takes $X_{L}=X_{R}$. In this case the allowed $X$ is
restricted \cite{He} either to $L_e-L_{\mu},\; L_e-L_{\tau}$ or
$L_{\tau}-L_{\mu}$. The current coupled to $U(1)_X$ boson $Z'$ is
vectorial when expressed in the weak basis in this case. Since
the initial choice of basis is arbitrary one could always
redefine the right-handed fields $e'_{iR}$ to obtain the choice
$X_L=X_R$. But the structure of physical current coupled to mass
eigenstates of fermions depends upon the choice of Higgs
fields. In the event of only one Higgs doublet neutral under
$U(1)_X$, the charged leptonic mass matrix is diagonal and the
physical current coupled to $Z'$ is vectorial. When one introduces
more Higgs fields transforming non-trivially under $U(1)_X$, the
$U(1)_X$ no-longer remains vectorial. The possible choices of
$X\equiv X_L=X_R$ are severely limited due to the anomaly
constraints, eq.(\ref{f1}).
In particular only three choices are possible which respectively
correspond to
\be
\barr{c} L_{\mu}-L_{\tau}\\L_e-L_\tau \\L_e-L_\mu\\ \earr
\hspace{0.5in} \barr{c} X={\hbox {diag}}(0,1,-1)\\
X={\hbox {diag}}(1,0,-1)\\
X={\hbox {diag}}(1,-1,0)\\ \earr \label{f2}
\ee
The structure of the current associated with the new $Z'$ can be
written as,
\be
{\cal L}_{Z'}={{g'} \over {\cos\theta}}
\left\{ {\bar e'}_{L} X \gamma_{\mu} e'_{L}+
{\bar e'}_{R} X \gamma_{\mu} e'_{R} \right\} Z'^\mu \label{c1}
\ee
where $e'_{LR}$ are column vector in generation space and
$\theta$ is the weak mixing angle introduced here purely for
notational convenience.
The coupling of the physical (i.e. mass eigenstate) fermions to $Z'$
depend upon the structure of the mass
matrix $M_l$ for the charged leptons. This is dictated by the
charge matrix $Q$ whose $(i,j)^{th}$ element correspond to the
$X$-charge of
bilinear ${\bar e'}_{iL} e'_{jR}$. For example, we have in case of
$L_e-L_{\tau}$

\be
Q=\left(\begin{array}{rrr}
              0&-1&-2\\
              1&0&-1\\
              2&1&0\\
              \end{array}\right) \label{Ql}
\ee
The possible structures of mass matrices follow from that of $Q$. In
particular, a Higgs field with charge $-Q_{ij}$ would contribute
to the $(i,j)^{th}$ element of the mass matrix $M_l$. Note that
the two different fields contribute to the $(M_l)_{ij}$ and
$(M_l)_{ji}$. Hence $M_l$ is necessarily
non-hermitian except when it is diagonal with only one Higg's
doublet carrying zero charge under $U(1)_X$.
In this case, the weak basis $e'_{L,R}$ coincide with the mass
basis $e_{L,R}$ and $Z'$ couples to a vector current
corresponding to $X$. When one introduces one or more additional
doublets transforming non-trivially
under $U(1)_X$ then $M_l$ is necessarily non-hermitian and can
be diagonalized by a biunitary transformation:
\bearr
      U_L M_l U^{\dagger}_R&=&{\hbox {diag}}.(m_e,m_{\mu},m_{\tau})
      \label{c2}\\
      e_{L,R}&=&U_{L,R} e'_{L,R} \label{c3}
\eearr
${\cal L}_ Z'$ then assumes the following form in terms of the
mass eigenstates:
\be
{\cal L}_{Z'}={{g'} \over {cos\theta}}(\kappa_{Lij} {\bar e}_{iL}
\gamma_{\mu} e_{jL}+
               \kappa _{Rij} {\bar e}_{iR}  \gamma_{\mu} e_{jR})
               Z'^{\mu} \label{c4}
\ee

where
\be
    \kappa_{a} \equiv {U_a X U^\dagger}_a \hspace{0.25in}   a=L,R.
    \label{c6}
\ee
Eq.(\ref{c4}) represents the general form of the $Z'$ interactions in
all the $SU(2)_L\otimes U(1)_Y \otimes U(1)_X$ models under study.
Different models are specified by
the choice of X and the Higgs fields which determine $M_l$ and hence
$U_{L,R}$.
The following two important properties are enforced by the structure
of X:

({\em i}) The current coupled to $Z'$ is non-vectorial except in a
specified
case $U_L=U_R=I$. This follows since $M_l$ is necessarily
non-hermitian when it
is not diagonal as already discussed. Hence $U_L \neq U_R$. Moreover,
for
$U_L \neq U_R$, $U_LXU^ \dagger_L$ and $U_RXU^\dagger_R$
cannot be identical\footnote{To prove this explicitly, we write
$U_{R}$ = $U_{L}V$ , $V$ being a unitary matrix different from
$I$. Then $U_{L}XU^{\dagger}_{L}$ = $U_{R}XU^{\dagger}_{R}$ only
if $VX$ = $XV$. This is not possible because of the restricted
structure of X.} leading to a non-vector current.

({\em ii}) The current coupled to $Z'$ would violate leptonic
flavor, i.e. $\kappa _{aij}$ are non-zero for $i \neq j$,
if $M_l$ is not diagonal. In
this case, $U_L$ and/or $U_R$ are different from unity. To see
this, consider $X\equiv L_e -L_\tau$. Because
of the form of $X$ given in eq.(\ref{f2}), it is easy to see that
$U_{L}XU^{\dagger}_{L}$ ( $U_{R}XU^{\dagger}_{R}$) will
have non-zero off diagonal couplings unless mixing between
$e'_L$($e'_R$) and $\tau '_L$($\tau '_R$) is forbidden. Since
such couplings would invariably occur in models with extended
Higgs structure, one expects the flavor changing $Z'$ couplings
in these cases.  This occurrence of the flavor changing current
is a well-known phenomena \cite{Weinberg} which arises
when fermions of the same charge and helicity transform differently
under a
gauge group; $U(1)_X$ in the present case.

Since the structure of the $Z'$ current is fixed by $X$ and $M_l$, it
 is
easy to classify all models that are possible within the present
scheme. One
has basically three types of models.

\noindent ({\em i}) Model with only one Higgs doublet neutral under
$U(1)_X$.
 In these,
eq.(\ref{Ql}) require $M_l$ to be diagonal. Hence one has vector
currents and no flavor
violation. There are three models in this category studied in
ref. \cite{He}.

\noindent ({\em ii}) Models with two Higgs carrying $U(1)_X$
charge 0,
and $\pm1$ or $ \pm2$. In
this case only one non-diagonal entry is possible in $M_l$
(see eq.(\ref{c5})). In
these types of models, one of the lepton remains unmixed while the
other two mix
with some mixing angles $\theta_{L,R}$.

\noindent ({\em iii}) Third category of the models follow when one
introduces
 two or more
additional Higgs fields carrying $U(1)_X$ charge $\pm1$ or $\pm2$.
These
represent general class of models with mixing among all three
generations.

We shall study in the next section detailed phenomenology of a model
in category ({\em ii}).

\section {$SU(2)_L \otimes U(1)_Y \otimes U(1)_{L_e-L_\tau}$ Model:}

\noindent We consider $SU(2)_L \otimes U(1)_Y \otimes U(1)_X$
model containing
standard fermions, two Higgs doublets $\phi_{1,2}$ and an $SU(2)_L
\otimes U(1)_Y$
singlet $\eta$. $X$ is chosen to be $L_e-L_\tau$ charge from
eq.(\ref{f2}). $X$ charges of
$\phi_1$ and $\phi_2$ are chosen to be 0 and +2 respectively.
The field $\eta$ is
assumed to carry some non-zero charge under $U(1)_X$ and it is solely
introduced to provide a
different mass scale characteristic of the $U(1)_X$ breaking.

The quark sector of the model remains the same as in the SM model
while lepton
couplings to the neutral Higgs fields are given by the following:
\bearr
-{\cal L}_Y&=&h_{ii} \bar e'_{iL}e'_{iR} \phi^{0}_1+
            h_{13} \bar e'_{1L}e'_{3R} \phi^{0}_2\\ \label{c12}
           & \equiv &{m_i \over \langle \phi^0_1 \rangle}
           \bar e'_{iL}e'_{iR} \phi^{0}_1+
           { \delta \over \langle \phi^0_2 \rangle} \bar
e'_{1L}e'_{3R}
\phi^{0}_2+h.c \non
\eearr

This leads to the following mass matrix $M_l$:
\be
M_l=\left(\begin{array}{ccc}
              m_1&0&\delta\\
              0&m_2&0\\
              0&0&m_3\\
              \end{array}\right)   \label{c5}
\ee

Let $U_{L,R}$ diagonalize $M_l$, i.e.
\[
     U_LM_lU^\dagger_R={\hbox {diag}}(m_e,m_\mu,m_\tau)
\]
where
\bearr
                 m_{\mu}^2&=&m_2^2  \non\\
m_e^2&=&{1 \over 2}\left\{ m_1^2+m^2_3+\delta^2+\left[
(m_1^2-m^2_3)^2+2\delta^2(m^2_1+m^2_3)+\delta^4\right]^{1 \over
2}\right\}  \non\\
m_{\tau}^2&=&{1 \over 2}\left\{ m_1^2+m^2_3+\delta^2-
\left[(m_1^2-m^2_3)^2+2\delta^2(m^2_1+m^2_3)+\delta^4\right]^{1 \over
2}\right\}\non
\eearr
\be
U_{L,R}=\left[\begin{array}{ccc}
              \cos\theta_{L,R}&0&\sin\theta_{L,R}\\
              0&1&0\\
              -\sin\theta_{L,R}&0&\cos\theta_{L,R}\\
              \end{array}\right] \label{nr1}
\ee
The mixing angles $\theta_{L,R}$ are given by:
\[
     \sin 2\theta_{L}=-{2\delta m_3 \over m_\tau^2-m^2_e}
\hspace{0.5in}
     \sin 2\theta_{R}=-{2\delta m_1 \over m_\tau^2-m^2_e} \label{c13}
\]
As we will soon see, the $\theta_{L,R}$ are constrained to be quite
small. It is therefore appropriate to work in the approximation
$\delta<m_1,m_3$. In this limit,
\be
\sin 2\theta_R \approx -{2 \delta m_e \over m_\tau^2}
\hspace{0.5in}
\sin 2\theta_L \approx -{2\delta \over m_\tau} \label{nr2}
\ee
The parameters $\kappa_{aij}$ ($a=L,R$) determining the couplings of
$Z'$ to leptons through eq.(\ref{c6}) are explicitly given in the
present case by
\bearr
\kappa_{a11}&=&\cos 2\theta_a=-\kappa_{a33}  \non\\
 \kappa_{a13}&=&-\sin 2\theta_a  \non\\
\kappa_{a2i}&=&0 \hspace {0.5in} i=1,2,3 \label{c7}
\eearr
Since one of the doublets carry non-zero $U(1)_X$ charge, the $Z'$
will mix with the conventional $Z$ boson to produce two mass
eigenstates $Z_{1,2}$.
\bearr
Z&=&\cos\phi \;Z_1+\sin\phi \;Z_2 \non\\
Z'&=&-\sin\phi \;Z_1+\cos\phi \;Z_2 \label{c8}
\eearr
The couplings of the neutral gauge boson $Z_{1,2}$ to the leptons
are now given by
\be
{\cal L}_{Z}={g \over \cos\theta}
\left\{ \sum_{m=1,2} F_{Lmij} {\bar e}_{iL}  \gamma_{\mu} e_{jL}
 Z^\mu_m+  L \leftrightarrow R \right\}
\ee
where
\bearr
F_{L1ij}&=&\cos\phi(-{1 \over 2}+\sin^2\theta)\delta_{ij}
-\sin\phi{g' \over g} \kappa_{Lij}\non\\
F_{R1ij}&=&\cos\phi \sin^2 \theta \delta_{ij}
-\sin\phi{g' \over g} \kappa_{Rij}\non\\
F_{L2ij}&=&\sin\phi(-{1 \over 2}+\sin^2\theta)\delta_{ij}
+\cos\phi{g' \over g} \kappa_{Lij}\non\\
F_{R2ij}&=&\sin\phi \;\sin^2\theta \delta_{ij}
+\cos \phi{g' \over g} \kappa_{Rij}\non
\eearr
As would be expected, eqs.(\ref{c5}) and (\ref{c7}) show that the
muon number is exactly conserved in the model. This is a consequence
of the fact that both the $Z'$ interactions as well as the mass
matrix, eq.(\ref{c5}), respect this symmetry.
When $\delta << m_{\tau}$, the flavor violations and departure from
vectorial symmetry are very small. Moreover, these departures are
more suppressed in the right-handed sector compared to the
left-handed sector.

   The generalization to other models in this category is
obvious. One could construct  another model with additional
Higgs carrying $L_{e}-L_{\tau}$ charge $-2$ instead of $+2$. In this
case $(M_{l})_{31}$ will be non-zero instead of $(M_{l})_{13}$
as in eq.(\ref{c5}). All the couplings of this model are then
obtained
by interchange of $\theta_{L} \leftrightarrow \theta_{R}$ in
eq.(\ref{c7}).
In addition to these two models with  $L_{e}-L_{\tau}$ symmetry,
one could construct pair of models each with symmetry
$L_{e}-L_{\mu}$ and  $L_{\mu}-L_{\tau}$. These are respectively
characterized
by an unbroken $L_{\tau}$ and $L_{e}$.

In addition to the flavor violations induced by $Z'$, there
exists other flavor violations associated with the Higgs fields.
These arise in a well-known \cite{Altarelli} manner whenever the
fermions with the same charge obtain their masses from two
different Higgses as in eq.(\ref{c12}). Using
eqns.(\ref{c12}-\ref{nr1}) it follows that
\[
-{\cal L}_{FCNC}=\delta\l({\phi_1^0 \o <\phi_1^0>}
-{\phi_2^0 \o <\phi_2^0>}\r)
\l{\cos\theta_L\sin\theta_R\bar{e_L}\tau_R +
\cos\theta_R\sin\theta_L\bar{\tau_L}e_R \r} +h.c.
\]
It follows from eq.(\ref{nr2}) that these flavor violations are
of $O\l(\delta^2/m_\tau <\phi_{1,2}^0>\r)$ and hence would be
suppressed in the limit $\delta << m_\tau$ compapred to $Z'$
induced flavor violations unless the associated Higgs is much
lighter than $Z'$. We shall therefore concentrate on the $Z'$
induced flavor violations in the next section.

We close this section with
a brief mention of the $Z$-$Z'$ mixing in these models and quote
well known formulae \cite{Altarelli} to be used later on. The neutral
gauge boson
mass matrix $M_{0}^{2}$ in the $Z$-$Z'$ basis is given by
\[
M_{0}^{2}=\left(\barr{cc}M_{Z}^{2} & \delta M^{2}\\
		         \delta M^{2} & M_{Z'}^{2}\\\earr\right)
\]
where
\[
M_{Z}^{2}=\frac{1}{4}
g^2 [\langle\phi_{1}\rangle^{2}+\langle\phi_{2}\rangle^{2}]
;\hspace{0.5in}
M_{Z'}^{2}=g'^{2} [\langle\phi_{2}\rangle^{2} +
\langle\eta\rangle^2]
\]
\be
\frac{\delta M^{2}}{M_{Z}^{2}} =4(g'/g)\sin^{2}\beta
\mbox{\hspace{0.25in}where\hspace{0.25in}}
\sin^{2}\beta=\frac{\langle\phi_{2}\rangle^{2}}{\langle\phi_{1}
\rangle^{2}+\langle\phi_{2}\rangle^{2}} \label{f3}
\ee
The mixing angle $\phi$ appearing in eq.(\ref{c8} ) is then given by
\be
\tan ^{2}\phi=\frac{M_{Z}^{2}-M_{1}^{2}}{M_{2}^{2}-M_{Z}^{2}}
\label{c11}
\ee
In addition, one has
\be
M_{1}^{2}\cos^{2}\phi +
M_{2}^{2}\sin^{2}\phi=\frac{M_{W}^{2}}{\cos^{2}\theta} \non\\
\hspace{0.25in}\hbox{and}\hspace{0.25in}
\sin\phi\cos\phi=\frac{\delta M^{2}}{M_{2}^{2}-M_{1}^{2}} \label{c9}
\ee
$\theta$($M_{W}$) being the Weinberg angle ($W$-mass) at the tree
level.
\section{Phenomenology of $SU(2)_L\otimes U(1)_Y \otimes
U(1)_{L_{e}-L_{\tau}}$}

We shall now explore the phenomenological consequences of the
$SU(2)_L\times U(1)_Y \times U(1)_{X}$ models. The extra $Z$-boson
associated with $U(1)_{X}$ change the phenomenology of the SM in
two ways. The extra $Z'$ contribute to the known processes
induced by the $Z$ boson. In addition, in the present case, $Z'$
induce new flavor violating processes. The detailed
phenomenology will depend upon the model. We shall take the
model presented in the last section as an illustrative example
and work out consequences within that model.

In the absence of additional Higgs, the $Z'$ induced flavor
violation disappears. Moreover the $Z'$ does not mix with the
ordinary $Z$. In this case $Z'$ makes its effect felt by
contributing to known processes like $e^{+}e^{-}\rightarrow
\mu^{+}\mu^{-}$ scattering. The detailed restrictions on the
relevant parameters by LEP results have been worked out in
ref. \cite{He} for this case. These restrictions continue to hold in
the
present case. But additionally one gets more stringent
restrictions due to flavor violations and $Z$-$Z'$ mixing. We
shall concentrate on these in the following.

The phenomenology of models with extra $Z$ boson is extensively
discussed in the literature \cite{Altarelli,Valle}. The present class
of models have characteristic
differences arising due to the fact that $Z'$ couples only to
leptons. In other models, an important restriction on the $Z'$
mass arises from the direct experimental observations at the
hadronic colliders. These restrictions though model dependent
strongly constrain the $Z'$ mass. For example in the left-right
symmetric model \cite{LR}, the search in $\bar{p}p$ collisions
imply \cite{PD}
$M_{Z_{LR}}>310$ GeV. Similar restrictions are not applicable here
since $Z'$ couples only to leptons. Its production at the
hadronic colliders arise only through mixing with the ordinary
$Z$ and is therefore highly suppressed. The $Z'$ mass as well as
its mixing with $Z$ is constrained in the present case by $(a)$
the observations at LEP and $(b)$ the observed limits on the
leptonic flavor violations. We discuss them in turn.
\subsection{Constraints from the LEP data}

We closely follow the analysis of ref. \cite{Altarelli} in deriving
constraints
on the relevant parameters from observations at LEP. These
constraints have been derived in two different ways. The
observations of the ratio $M_{W}/M_{1}$ and the $Z$-mass
$M_{1}$, at CDF and LEP respectively, constrain the $\rho$ parameter
and lead to restrictions on $M_{2}$ and $\tan\phi$. Other method
is to use the fact that the extra $Z$ induce changes in
observables like width to fermions, peak cross section in
$e^+e^-$ collisions etc. One
could then make a detailed fit to the LEP data and derive
constraints on $M_{2}$ and $\phi$.

The mixing between $Z$ and $Z'$ change
the tree level relation between the $W$ and the $Z$ mass.
Specifically,
\[
\frac{M_{W}^{2}}{\rho_M M_{1}^{2}}=\cos^{2}\theta
\]
$\theta$ being the tree level weak mixing angle. The parameter
$\rho_M$ can
be read off from the mixing matrix between $Z$ and $Z'$ (see
eq.(\ref{c9})):
\be
\rho_{M}=\frac{1+\tan^2\phi\frac{M^2_2}{M^2_1}}{1+\tan^2\phi}
\label{f7}
\ee
One could eliminate $\cos^{2}\theta$ in favor of $G_{F}$,
$\alpha$ and $M_{1}$ to obtain
\be
\frac{M_{W}^{2}}{\rho_M M_{1}^{2}}=
\left(\frac{1}{2}+\sqrt{\frac{1}{4}-\frac{\mu^{2}}{\rho_M
M_{1}^{2}}}\right)\label{c10}
\ee
\hbox{where}
\[
\mu=\sqrt{\frac{\pi\alpha}{\sqrt{2}G_{F}}}=(37.280\hbox{GeV})
\]
These restrictions are valid at the tree level. Since the extra
$Z$ induced effects are comparable to the radiative corrections
in the standard model, one must incorporate the later. This has been
done in

ref. \cite{Altarelli}, assuming that the radiative corrections
induced by
$Z_{2}$ are negligible. The radiative corrections of SM are
included using the improved Born approximation which changes
eq.(\ref{c10}) to the following:
\be
\frac{M_{W}^{2}}{\rho M_{1}^{2}}=
\left(\frac{1}{2}+\sqrt{\frac{1}{4}-\frac{\mu^{2}}{\rho
M_{1}^{2}(1-\Delta\alpha)}}\right)\label{r1}
\ee
where the $\rho$ parameter is now
\[
\rho=\frac{\rho_{M}}{1-\Delta\rho_{T}}
\]
with
\[
\Delta\rho_{T}\simeq 3\frac{G_{F}m_{t}^{2}}{8\pi^{2}\sqrt{2}}
\hspace{0.25in} \hbox{and} \hspace{0.25in}
\Delta\alpha =0.0602+\frac{40}{9}\frac{\alpha}{\pi}\ln
\frac{M_{1}{\hbox {(GeV)}}}{92} \pm 0.0009
\]
The CDF result on $\frac{M_{W}}{M_{1}}=0.779$ together with the
LEP result on\footnote{Note that unlike, the fermionic width, the
determination of
$M_{1}$ from the data is fairly insensitive to the presence of $Z'$.}
the $Z$-mass $M_{1}$ can be used to obtain
$\rho=1.005\pm 0.003$ in eq.(\ref{r1}). This implies at
$1\sigma $
\be
\Delta\rho_{M}=\rho_{M}-1\leq
0.008-0.003\left(\frac{m_{t}{\hbox {(GeV)}}}{100}\right)^{2}
\label{f6}
\ee

In addition to this restriction, $\Delta\rho_M$ can also be
constrained \cite{Altarelli,Valle} by the other observables at LEP.
Specifically, the presence of $Z'$ would change the three
leptonic widths $\Gamma_{e\, \mu\, \tau}$ as well as the hadronic
width $\Gamma_h$ of the $Z_1$. These changes can be parameterized
\cite{Altarelli}
in terms of $\Delta\rho_M$ and mixing angle $\phi$:
\be
d\Gamma_i = A_i \Delta\rho_M + B_i \phi \label{f4}
\ee
In our case
\bearr
A_i&=&4N_{c}\rho_f\left[(T_{3Li}-\sin^2\theta_fQ_i)^2 + T_{3Li}^2 +
      {4\sin^2\theta_f\cos^2\theta_f \over \cos
2\theta_f}Q_i(T_{3Li}-\sin^2\theta_fQ_i)\right]\non \\
B_i&=&8N_{c}\rho_f\left[(T_{3Li}-\sin^2\theta_fQ_i)g'_{Vi}-T_{3Li}
g'_{Ai}\right]\non
\eearr
where
\[
  g'_{Vi}={g'\over g}(\kappa_{Lii}+\kappa_{Rii}) ; \;\;\;
            g'_{Ai}={g'\over g}(\kappa_{Lii}-\kappa_{Rii}) ;
\]
\[
 \rho_f\equiv {\rho\over\rho_M} ;\;\;\; \sin^2\theta_f=
{1\o 2}-\sqrt{{1\o 4}-{\mu^2\o {\rho_fM_1^2(1-\Delta_\alpha)}}}
\]
$N_{c}=3(1+{\alpha_s\o \pi})$ for quarks and 1 for leptons.
Fermionic width $\Gamma_i$ of $Z_1$ have been extracted from the
LEP data in a model independent way. We use the values derived
in ref. \cite{Gurtu}
 to constrain $\Delta\rho_M$ and $\phi$. Specifically,
\[
\barr{rcl}
\Gamma_e&=&82.6\pm 0.7{\hbox {MeV}}\\
\Gamma_\mu&=&83.6\pm 1.1{\hbox {MeV}}\\
\Gamma_\tau&=&83.1\pm 1.2{\hbox {MeV}}
\earr
\hspace{0.5in}
\Gamma_h=1.741\pm 0.015{\hbox {MeV}}
\]
We use these values and determine the best values for
$\Delta\rho_M$ and $\phi$ appearing in eq.(\ref{f4}) through a
least square fit. This gives (for $m_t=150$GeV) at 1$\sigma$:

\be
\Delta\rho_M=-0.0018\pm 0.004\, \hspace{0.5in}
\phi=0.0094\pm 0.012 \label{f5}
\ee
The value of $\Delta\rho_M$ as determined by eq.(\ref{f5}) is less
stringent than following from eq.(\ref{f6}) derived on the basis of
the
CDF result on ${M_W \over M_1}$. We shall therefore use the
values given by eq.(\ref{f6}) for $\Delta\rho_M$ in the next section
to constrain the
parameters of the model.

\subsection{Constraints from the rare processes}

As already discussed, the model of the last section contains
flavor violations involving $\tau\; \hbox {and}\; e$. The muon number
is
exactly conserved in the model. As a consequence one expects the
following rare processes to occur in the model:\nopagebreak
\begin{itemize}
\item $Z_{1,2}\rightarrow e\tau$\samepage
\item $\tau\rightarrow eee$
\item $\tau\rightarrow e\mu\mu$
\end{itemize}
The branching ratios for these processes can be easily worked
out and are given by:
\bearr
\frac{\Gamma (\tau\rightarrow eee)}{\Gamma (\tau\rightarrow
\nu_{\tau}\nu_{e}e)}
& = &
16M_{1}^{4}\left\{(g^{e}_{LL})^{2}+(g^{e}_{RR})^{2}+\frac{1}{2}
\left((g^{e}_{LR})^{2}+(g^{e}_{RL})^{2}\right)\right\}\non\\
\frac{\Gamma (\tau\rightarrow e\mu\mu)}{\Gamma (\tau\rightarrow
\nu_{\tau}\nu_{e}e)}
& = &
4M_{1}^{4}\{(g^{\mu}_{LL})^{2}+(g^{\mu}_{RR})^{2}+(g^{\mu}_{LR})^{2}+
(g^{\mu}_{RL})^{2}\}
\non\\
\Gamma (Z\rightarrow\tau e) & = & \frac{G_{F}M_{1}^{3}}{3\sqrt
{2}\pi}\{(F_{L1}^{\tau e})^{2} + (F_{R1}^{\tau e})^{2}\} \non
\eearr
where
\[
g_{LL}^{m}=\frac{F_{L1}^{\tau e}F_{L1}^{mm}}{M_{1}^{2}} +
\frac{F_{L2}^{\tau e}F_{L2}^{mm}}{M_{2}^{2}}\\
\hspace{0.5in}
g_{LR}^{m}=\frac{F_{L1}^{\tau e}F_{R1}^{mm}}{M_{1}^{2}} +
\frac{F_{L2}^{\tau e}F_{R2}^{mm}}{M_{2}^{2}}
\]
$m=e,\mu$. $g_{RR}$ and $g_{RL}$ are obtained by
$L\leftrightarrow R$ interchange in above equation.
The difference in the rates for the $\tau\rightarrow eee$ and
$\tau\rightarrow e\mu\mu$ arise due to both the $s$ and $t$ channel
$Z_{1,2}$ exchanges contributing to the former. In addition to
constraints
from the LEP discussed earlier the rare decays also provide
important constraints on the model. The specific constraints are
\cite{PD} given by the following:
\[
\barr{lcl}
Br (Z\rightarrow e^{+}\mu^{-}) & < & 2.4\times 10^{-5}\\
Br (Z\rightarrow e^{+}\tau^{-}) & < & 3.4\times 10^{-5}\\
Br (Z\rightarrow \mu^{+}\tau^{-}) & < & 4.8\times 10^{-5}\\
Br (\tau\rightarrow eee) & < & 2.7\times 10^{-5}\\
Br (\tau\rightarrow e\mu\mu) & < & 2.7\times 10^{-5}\\
Br (\tau\rightarrow \mu\mu\mu) & < & 1.7\times 10^{-5}\\
\earr
\]

The basic parameters of models are mixing angles $\theta_{L,R}$,
$Z_{2}$ mass $M_{2}$, $Z$-$Z'$ mixing angle $\phi$ and the
$U(1)_X$ gauge coupling $g'$. Both the $Z$-$Z'$ mixing and the
flavor violation arise in the model from the presence of the
additional doublet $\phi_{2}$. Thus both are related to the
parameter $\tan\beta=\langle\phi_{2}\rangle /\langle\phi_{1}\rangle$.
Relation
between $\phi$ and $\beta$ follows from eq.(\ref{f3}) and (\ref{c9})
\be
\sin\phi\sim 4C\left(\frac{M_{1}}{M_{2}}\right)^{2}\sin^{2}\beta
\label{c14}
\ee
where
\[
C\approx\frac{g'}{g}\left(1-\frac{M_{1}^{2}}{M_{2}^{2}}\right)^{-1}
\sim O(1)
\]
The $\theta_{L,R}$ also goes to zero when $\beta\rightarrow 0$.
If one assumes that the flavor violating Yukawa coupling
$h_{13}$ in eq.(\ref{c12}) is of the same order as flavor conserving
one
(namely $h_{33}$) then $\delta\approx m_{\tau}\tan\beta$ and hence
from eq.(\ref{c13})
\be
\sin 2\theta_{L}\approx -2\tan\beta \label{c15}
\ee

The existing limits on the $Br(\tau\rightarrow eee)$ as well as
$Z\rightarrow e\tau$ imply restrictions on the parameters $\beta$
and $M_2$. These are displayed in fig.1 assuming $h_{13}=h_{33}$.
Analogous constraints also follow from the process
$\tau\rightarrow e\mu \mu $. This process is comparatively
suppressed in the present case and hence imply much weaker
constraints. This is not displayed in the figure for
simplicity. The same parameters are also constrained by
$\Delta\rho_M$ and $\phi$( see eqs.(\ref{f7}) and (\ref{c14})).

It follows that the strongest
constraints on the parameters are implied by the rare decay
$\tau\rightarrow eee$. Hence the process $\tau\rightarrow eee$
is allowed by the LEP data to occur at a rate consistent with
the present experimental precision. Improvement in the limits
for this process would either imply more stringent restrictions
on $\beta$ and ${\hbox {M}}_2$ or one should be able to see this
decay in
future. Fig.1 was based on the assumption of equal Yukawa
couplings, $h_{13}=h_{33}$, in eq.(\ref{c12}). For comparison we also
display in fig.2 limits on $\beta$ and ${\hbox {M}}_2$ in case of
$h_{13}=10^{-2}h_{33}$. Reduction in the value of $h_{13}$ strongly
suppresses the flavor violating couplings of $\tau$.
$\Delta\rho_M$ and $\phi$ remain unchanged. As a result, now the
LEP data imply stronger restrictions on $\tan\beta$ and ${\hbox
{M}}_2$. In this case, the LEP observations already rule out
possibility of seeing flavor violation in future experiments
which are expected to provide improved limits on $\tau\rightarrow
eee$.

It is clear from fig.1 and 2 that as long as ${\hbox {M}}_2 <
O(TeV)$, $\tan\beta$ is restricted to be $< O(0.1-0.5)$. Hence
the vacuum expectation value of the field $\phi_2$ responsible
for flavor violations is strongly constrained in the model.
Likewise, low values of $M_2$ (e.g. 400GeV) are possible only if
$\tan\beta$ is chosen small (0.03 in case of $h_{13}=h_{33}$, and
0.3 in case of $h_{13}=10^{-2}h_{33}$).

Although we restricted ourselves to the $L_e-L_\tau$ model, the
analogous
constraints would follow in models with $X=L_e-L_\mu$ or
$L_\mu-L_\tau$. In particular, one would expect very severe
constraint if $L_e-L_\mu$ is gauged since $\mu\rightarrow eee$
is much severely constrained experimentally.

\section{Summary}
We have studied in this paper a specific class of extended gauge
models of the form \linebreak $SU(2)_L\otimes U(1)_Y\otimes U(1)_X$.
All
these extensions are characterized by the fact that it is
possible to gauge $U(1)_X$ without extending the fermionic
sector of the standard model. Thus models studied here are the
simplest gauge extensions of the SM. These models are prototype
of more general horizontal symmetries \cite{Weinberg}. We have
concentrated here ({\em a}) on a systematic classification of
$U(1)_X$ models
and ({\em b}) on deriving constraints on parameters of a
prototype model using the LEP results. In specific case of $U(1)_X$
coupling to leptons, we have
categorized all possible choice of $U(1)_X$. In general
$U(1)_X$ provide important restrictions on the mixing
matrices. Moreover, they also give rise to interesting flavor
violations thus providing window into the existence of such
symmetry. The mixing of the $U(1)_X$ gauge boson
$Z'$ with the ordinary $Z$ is correlated in these models to the
flavor violation. In fact both these features originate from the
existence of a Higgs doublet carrying non-zero $U(1)_X$ charge.
As a result the observations at LEP could indirectly provide
important constraints on flavor violations. Detailed study
presented here shows that under reasonable assumptions on
relevant Yukawa couplings, the LEP observations do allow
sizeable flavor violations and it is possible to obtain rate
for $\tau\rightarrow eee$ near its present experimental limit.
In contrast, the lepton flavor violating decays of $Z$ are
considerably suppressed in these models.

We mainly studied models in which $U(1)_X$ acts only on leptons.
Models with $U(1)_X$ acting on quarks \cite{HM} or both can be analogously
studied. A systematic study of these horizontal models and
restrictions on flavor violations in these models in the light
of LEP observations would be interesting in its own right.

\newpage
\begin{center}
{\large{\bf Figure Caption}}
\end{center}
{\bf Figure 1:} The allowed region in the $M_2$-$\tan\beta$ plane
implied by various constraints:\\
 Curve(A) is a contour for
$Br(\tau\rightarrow eee)=2.7\times 10^{-5}$; (B) for
$Br(Z\rightarrow e\tau)=3.4\times 10^{-5}$; (C) for
$\Delta\rho_M=0.00125$;
and (D) for $\phi=0.021$. These curves are for $h_{13}=h_{33}$ (see
text). Region to the left of the curves is allowed.\\
{\bf Figure 2:} Same as figure 1 except that $h_{13}=10^{-2}h_{33}$.

\end{document}